\documentclass[aps,prl,reprint,superscriptaddress,amsmath,amssymb,showpacs]{revtex4-1}

\usepackage{appendix}
\usepackage{graphicx}
\usepackage[usenames]{color}

\begin{document}

\title{Imaginary time, shredded propagator method for large-scale GW calculations}

\author{Minjung Kim}
\affiliation{
Department of Applied Physics, Yale University,
New Haven, Connecticut 06520, USA
}
\author{Glenn J. Martyna}
\affiliation{
IBM TJ Watson Laboratory, Yorktown Heights, New York, USA 
}

\author{Sohrab Ismail-Beigi}
\email{sohrab.ismail-beigi@yale.edu}
\affiliation{
Department of Applied Physics, Yale University,
New Haven, Connecticut 06520, USA 
}

\date{\today}

\begin{abstract}
The GW method is a many-body approach capable of providing quasiparticle bands for realistic  systems spanning physics, chemistry, and materials science.  Despite its power, GW is not routinely applied to large complex materials due to its computational expense. We perform an exact recasting of the GW polarizability and the self-energy as Laplace integrals over imaginary time propagators. We then ``shred'' the propagators (via energy windowing) and approximate them in a controlled manner by using Gauss-Laguerre quadrature and discrete variable methods to treat the imaginary time propagators in real space. The resulting cubic scaling GW method has a sufficiently small prefactor to outperform standard quartic scaling methods on small systems ($\gtrapprox$ 10 atoms) and also represents a substantial improvement over several other cubic methods tested. This approach is useful for evaluating quantum mechanical response function involving large sums containing energy (difference) denominators. 
\end{abstract}

%\pacs{73.40.-c, 68.35.-p, 68.35.Rh}
\maketitle

Density Functional Theory (DFT)~\cite{hohenberg_inhomogeneous_1964,kohn_self-consistent_1965} within the local density (LDA) or generalized gradient (GGA)~\cite{perdew_self-interaction_1981,perdew_atoms_1992} approximation provides a solid workhorse capable of realistically modeling an ever increasing number and variety of physical systems spanning condensed matter, chemistry, and biology.  Generally, this approach provides a highly satisfactory description of the total energy, electron density,  atomic geometries, vibrational modes, etc. However, DFT is a ground-state theory for electrons and DFT band energies do not have direct physical meaning (DFT is not a quasiparticle theory).  In addition, there are significant failures when DFT band structures are used to predict electronic excitations \cite{perdew_density-functional_1982, lundqvist_theory_2013, 
anisimov_first-principles_1997}.

The GW approximation to the electron self-energy \cite{hedin_new_1965,hybertsen_electron_1986, aryasetiawan_gw_1998,onida_electronic_2002} is one of the most accurate fully {\it ab initio} methods for the prediction of electronic band structures which can be used to correct the approximate DFT results.  Despite its power, GW is not routinely applied to complex materials systems due to its unfavorable computational scaling: the cost of a standard GW calculation scales as $O(N^4)$ where $N$ is the number of atoms in the simulation cell whereas Kohn-Sham DFT calculations scale as $O(N^3)$.  

Hence, reducing the expense of GW calculations has been the subject of numerous  studies.  $O(N^4)$ GW methods with  smaller prefactors  avoid the use of unoccupied states via iterative matrix inversion~\cite{wilson_efficient_2008, wilson_iterative_2009, rocca_ab_2010,lu_dielectric_2008, giustino_gw_2010,umari_gw_2010,govoni_large_2015} or use sum rules or energy integration to greatly reduce the number of unoccupied states~\cite{bruneval_accurate_2008,berger_ab_2010,gao_speeding_2016}. 
Prior cubic-scaling $O(N^3)$ methods include a spectral representation approach~\cite{foerster_on3_2011} (which has a large prefactor as we discuss below) and a space/imaginary time method~\cite{liu_cubic_2016} requiring analytical continuation from imaginary to real frequencies.
Finally, even linear scaling GW is possible via stochastic approaches~\cite{neuhauser_breaking_2014} for the total density of electronic states with the caveat that the non-deterministic stochastic noise must be added to the list of usual convergence parameters.

Here, we present a deterministic $O(N^3)$ GW approach in real space based on sum-over-states arising from a imaginary time formulation which forms the basis for  controlled approximations. The method shows excellent convergence by using an exact energy windowed Laplace transform over imaginary time allowing for accurate treatment with Gauss-Laguerre quadrature integration.  As we show below, the windowing strategy leads to very efficient reduced order method with a small prefactor, 
and hence our $O(N^3)$ method is already competitive with both $O(N^4)$ approaches for small unit cells (and is thus guaranteed to win for even larger systems). Similarly, it outperforms other $O(N^3)$ methods we have  tested. As our approach works directly in frequency domain bypassing imaginary time samplings or analytic continuation~\cite{liu_cubic_2016}, it is easy to implement in standard GW implementations~\cite{deslippe_berkeleygw:_2012,supplement}. 

To keep the discussion simple, we describe how the new approach works for the basic and most widely used ``G$_0$W$_0$'' level of GW theory: both the screening and self-energy are computed based on the DFT band structure with no further self-consistency. 
Our approach is directly applicable to G$_0$W$_0$, while applying it to more complex GW calculations requires further developments.
For clarity, we develop our method using the static random phase approximation irreducible polarizability matrix $P$ for an insulating system with an energy gap to demonstrate the basic principles.  The modifications needed to handle finite temperature, metals the self-energy are described afterwards (and in Ref.~\cite{supplement}).

In real space, for a zero temperature gapped system  we have 
\begin{equation}
P_{r,r'} = -2\sum_{v}^{N_v} \sum_{c}^{N_c} \frac{\psi_{r,v}^*\psi_{r,c}\psi_{r',c}^*\psi_{r',v}}{E_c-E_v}
\label{eq:P}
\end{equation}
where $N_v$ and $N_c$ are the number of occupied (valence  $v$) and unoccupied (conduction  $c$) states.  The single particle states have real-space wave function values $\psi_{r,n}=\psi_n(r)$ and energies $E_n$. For clarity, we suppress non-essential quantum numbers such as spin $\sigma$ and Bloch $k$-vectors.  (Spin is simply tacked onto $r$ via $r\rightarrow(r,\sigma)$; including crystal momentum requires these replacements: $P_{r,r'}\rightarrow P^q_{r,r'}$ where $q$ is  momentum transfer,  $\psi_{r,v}\rightarrow\psi_{r,vk}$,  $E_v\rightarrow E_{vk}$,  $\psi_{r,c}\rightarrow\psi_{r,ck+q}$,  $E_c\rightarrow E_{ck+q}$, sum Eq.~(\ref{eq:P}) over $k$ and divide by the number of k-points.)  Current numerical methods for computing $P$ based on the sum-over-states formula of Eq.~(\ref{eq:P}) have an $O(N^4)$ scaling (e.g.,  Ref.~\cite{deslippe_berkeleygw:_2012}).  Approaches that reduce the expense of computing $P$, the most computational intensive part of GW, are welcome.

The key advantage of working in the real space representation of $P$ (Eq.~\ref{eq:P}) is that the product over wave functions is already separable: if the energy dependence (i.e., the energy denominator) can be made separable, one can reduce the algorithmic scaling by an order to $O(N^3)$.  %A simple approach based on interpolating the energy dependence in the sum for $P$ is detailed in~\cite{supplement} and its performance is assessed below.  
We describe a method based on Laplace transforms over imaginary time and subsequent numerical approximation by Gaussian quadrature that delivers high performance (see Ref.~\cite{supplement} for an alternative but less efficient approach based on interpolation). 

Since $E_c-E_v\ge E_{g}>0$, where $E_g$ is the energy gap, the Laplace transform  
\begin{equation}
\frac{1}{E_c-E_v} = \frac{1}{a}\int_0^\infty dx\ e^{-x(E_c-E_v)/a}
\label{eq:laplace}
\end{equation}
neatly creates the desired separability under the integral; here $a$ is an energy scale parameter discussed below. 
%The imaginary time $\tau=x/a$ connects our method to other imaginary time GW methods~\cite{rieger_gw_1999,kaltak_low_2014,liu_cubic_2016}.  
Inserting Eq.~(\ref{eq:laplace}) into Eq.~(\ref{eq:P}) leads to the separable form
\begin{equation}
P_{r,r'} = -\frac{2}{a} \int_0^\infty dx\ e^{-xE_{g}/a}\,\bar\rho_{r,r'}(x/a)\,\rho_{r',r}(x/a)
\label{eq:Psepexact}
\end{equation}
where
\begin{eqnarray}
\bar\rho_{r,r'}(\tau) & = & \sum_c^{N_c} e^{-\tau\Delta E_c}\psi_{r,c}\psi_{r',c}^*\,,
\label{eq:rhorhobardef1}\\
\rho_{r,r'}(\tau) & = & \sum_v^{N_v} e^{-\tau\Delta E_v}\psi_{r,v}\psi_{r',v}^*\,,
\label{eq:rhorhobardef2}
\end{eqnarray}
are the unoccupied and occupied imaginary time propagators (Green's functions), respectively. We have introduced the valence band maximum $E_{v}^{max}$, conduction band minimum $E_c^{min}$ and band gap $E_g=E_c^{min}-E_{v}^{max}$ to ensure we have decaying exponentials with increasing energy away from band edges by defining $\Delta E_v \equiv E_v^{max}-E_v$ and $\Delta E_c \equiv E_c - E_c^{max}$. The imaginary time formalism connects our work to that of Ref.~\cite{rieger_gw_1999,kaltak_low_2014,liu_cubic_2016}.  

Formally, the exact formula Eq.~(\ref{eq:Psepexact}) represents an $O(N^3)$ method for systems represented in a finite basis set scaling with $N$, since the sums over $v$ and $c$  are separable.  In practice, the integral over imaginary time $x/a$ must be replaced by a discrete quadrature.  If a quadrature is applied directly to Eq.~(\ref{eq:Psepexact}) without further refinement, to achieve tolerable errors quadrature grid must be taken to be very fine.  The reason is straightforward: for a well-converged GW calculation, many high energy conduction bands are needed so that the energy differences $E_c-E_v$ becomes quite large leading to rapidly decaying exponentials in $x$ which necessitates dense quadrature grids in $x$.  More precisely, the smallest and largest energy scales are the gap  $E_g=E_{c}^{min}-E_v^{max}$ and the bandwidth $E_{bw}=E_c^{max}-E_v^{min}$, and $E_{bw}/E_g>100$ is typical especially for small-gapped materials.

\begin{figure}
\includegraphics[width=2.5in]{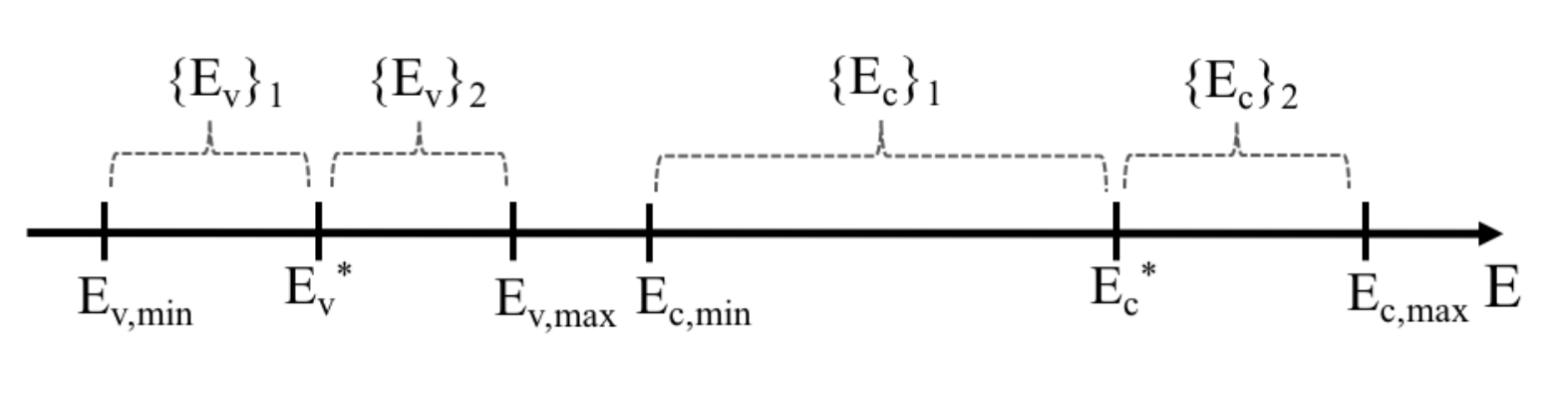}
\caption{Example of $2\times2$ windowing with two valence and two conduction windows $N_{vw}=N_{cw}=2$. $E_v^*$ and $E_c^*$ are the energy points dividing the valence and conduction windows.}
\label{fig:2x2wandcost}
\end{figure}

To alleviate the large bandwidth/small gap problem, we introduce an exact energy windowing approach based on  shredding (decomposing) the  propagators: we divide the energy range of the valence band into $N_{vw}$ contiguous energy windows and similarly for $N_{cw}$ conduction band windows. Valence window $l$ ranges from $E_{vl}^{min}$ to $E_{vl}^{max}$ (conduction band windows are indexed by $m$). Figure~\ref{fig:2x2wandcost} shows a simple example of a 2$\times$2 window decomposition.  This exact rewriting simply regroups the band summations into batches over pairs of energy windows:
\begin{equation}
P_{r,r'} = \sum_l^{N_{vw}} \sum_m^{N_{cw}} P^{lm}_{r,r'}
\label{eq:Pwindowed}
\end{equation}
where each window pair $(l,m)$ contributes
\begin{equation*}
P^{lm}_{r,r'} = -\frac{2}{a_{lm}} \int_{0}^\infty \!\!\!dx\, e^{-xE_{g}^{lm}/a_{lm}} \bar\rho_{r,r'}^{m}(x/a_{lm})\rho_{r',r}^{l}(x/a_{lm}).
\end{equation*}
Note, each window pair has its own energy range $a_{lm}$ and the imaginary time density matrices for the windows are given by
\begin{eqnarray}
\bar\rho_{r,r'}^m(\tau) & = & \sum_{c\in m} e^{-\tau\Delta E_{cm}}\psi_{r,c}\psi_{r',c}^*\,,
\label{eq:rhorhobarlm1}\\
\rho_{r,r'}^l(\tau) & = & \sum_{v
\in l} e^{-\tau\Delta E_{vl}}\psi_{r,v}\psi_{r',v}^*\,.
\label{eq:rhorhobarlm2}
\end{eqnarray}
and $\Delta E_{vl}=E_{vl}^{max}-E_v$ and $\Delta E_{cm}=E_c-E_{cm}^{min}$ defined with respect to the extreme band energies in each window.  A good choice of windows can significantly reduce the ratio $E^{lm}_{bw}/E^{lm}_{g}$ for a window pair which allows the use of coarse quadrature grids and hence an efficient method.

While Eq.~(\ref{eq:Pwindowed}) is exact as written, we need to calculate it accurately via controlled, efficient approximations.  First, we must discretize the $r$-coordinate to generate finite-sized matrices.  For the widely used plane wave Fourier basis, which we employ herein, we use a uniform grid in $r$-space that is dual to the finite Fourier ($g$-space) basis; one combines this with  fast Fourier transforms (FFTs) to move between the Fourier and $r$ representations exactly.  
For other basis sets, appropriate real-space  discrete variable representations (DVRs) can be used \cite{baye_heenen_1986,friesner_solution_1986,dvrbook}. Second, the imaginary time integrals must be discretized, which is what we focus on below.  Given the exponentials being integrated, we use Gauss-Laguerre (GL) quadrature with $N_{GL}$ points: 
\begin{equation}
\int_0^\infty dx\ e^{-x}f(x) \approx \sum_{k=1}^{N_{GL}} w_k\,f(x_k)
\label{eq:GLquad}
\end{equation}
where $\{w_k\}$ and $\{x_k\}$ are weights and nodes for GL quadrature~\cite{abramowitz_handbook_1972} whose $N_{GL}$ dependence has been suppressed for clarity.  The contribution from window pair $(l,m)$ to Eq.~(\ref{eq:Pwindowed}) is  approximated by
\begin{equation}
P^{lm}_{r,r'} = -\frac{2}{a_{lm}}\sum_{k=1}^{N_{GL}^{lm}} w_k e^{-x_k(E_{g}^{lm}/a_{lm}-1)} \bar\rho_{r,r'}^{m}(x_k/a_{lm})\rho_{r',r}^{l}(x_k/a_{lm})\,.
\end{equation}

Choosing the energy scale $a_{lm}$ is a straightforward matter of minimizing errors~\cite{supplement}:  $a\approx\sqrt{E_{g}^{lm}E_{bw}^{lm}}$ is very close to the optimal choice. To quantify $N_{GL}$, we consider the target function $\hat P$ where all $\psi_n(r)=1$,
\begin{equation}
\hat P = \sum_c\sum_v\frac{1}{E_c-E_v}\,.
\label{eq:Phat}
\end{equation}
We then repeat the Laplace transform, windowing and quadrature steps for $\hat P$.
Assuming a flat density of states for valence and conduction bands, the errors in GL quadrature of $\hat P$ for each window pair turn out to depend primarily on $E_{bw}^{lm}/E_{g}^{lm}$ for that window pair. For a fixed error tolerance, we find that $N_{GL}^{lm}\propto\sqrt{E_{bw}^{lm}/E_{g}^{lm}}$~\cite{supplement}.  For a material such as Si where the DFT $E_{g}\approx 0.5$ eV and $E_{bw}\approx55$ eV is needed for good convergence, not using any windows translates into $N_{GL}\approx 20$ which is large; windowing is the remedy.

The final step is to choose an optimal windowing that minimizes the overall computational cost.
The cost to compute $P_{lm}$  scales as $N_{GL}^{lm}(N_c^{lm}+N_v^{lm})$.  Assuming flat densities of states $D_v$ and $D_c$ for the valence and conduction bands, respectively, where $D_v=N_v/(E_{v}^{max}-E_v^{min})$ (and similarly for $D_c$), we have that 
$N_v^{lm} =  ( E_{vl}^{max}-E_{vl}^{min})D_v$ (and similarly for $N_c^{lm}$).  Altogether, the total computational cost $C$ of evaluating $P$ is
\begin{equation}
C\! \propto\!\! \sum_l^{N_{vw}}\!\!\sum_m^{N_{cw}} \sqrt{\frac{E^{lm}_{bw}}{E^{lm}_{g}}} 
 \left[ \frac{E_{vl}^{max}\!-\!E_{vl}^{min}}{E_v^{max}\! - \!E_v^{min}}N_v \!+ \!
\frac{E_{cm}^{max}\!-\!E_{cm}^{min}}{E_c^{max}\! -\! E_c^{min}}N_c
\right]\,.
\label{eq:cost}
\end{equation}
This expression for $C$ compares very well to a more explicit evaluation of $C$ using actual values of $N_{GL}^{lm}$ and sums over the transition energies in the windows, which can alternatively be employed to optimize for systems with complicated DOS structure~\cite{supplement}.

In principle, we should minimize Eq.~(\ref{eq:cost}) over all possible number of windows and positions of the window boundaries.  {\it A posteriori}, this is unnecessary given   the smooth behavior of $C$:  simpler approaches are equally effective.  We   vary the number of windows $N_{cw}$ and $N_{vw}$ from 1 to 10 independently, and the window boundaries are always chosen to be from a list of fixed list of energies that divide each band into 10 equal segments.  For a given number of windows $(N_{vw},N_{cw})$, we minimize the cost function of Eq.~(\ref{eq:cost}) over all the discrete window choices. For example, to simulate bulk Si with its relatively small gap of $E_g=0.5$ eV, when $E_{bw}=54.5$ eV, the minimum number of computation occurs at $N_{vw}=1$ and $N_{cw}=4$~\cite{supplement}.

% Figure - epsilon inverse
\begin{figure}
\includegraphics[width=2.4in]{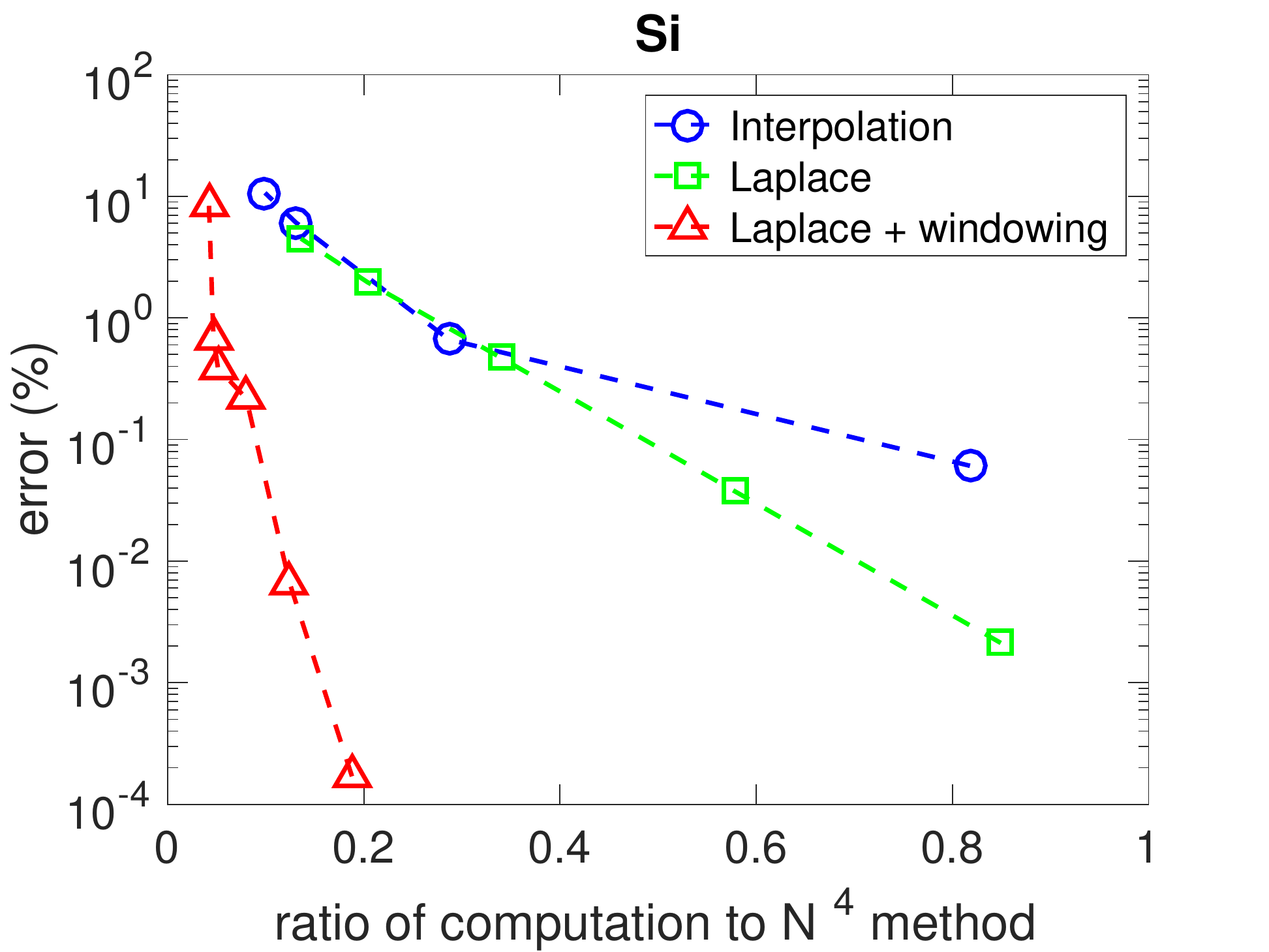}
\includegraphics[width=2.4in]{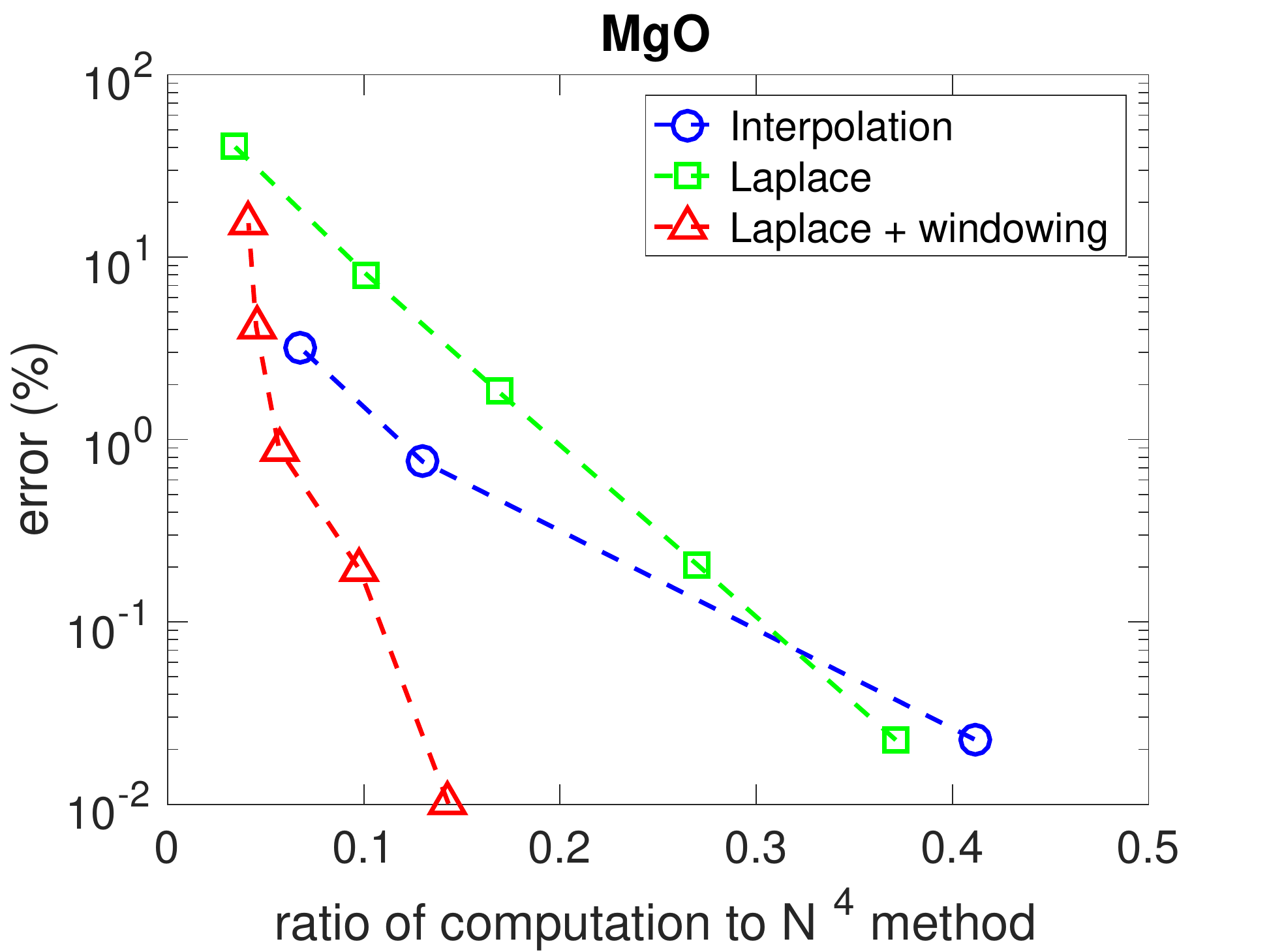}
\caption{Error in the macroscopic RPA optical dielectric constant $\epsilon_{\infty}$ for the interpolation, the naive Laplace GL, and the windowed Laplace GL methods with respect to the quartic $O(N^4)$ method.  The horizontal axis is the ratio of computational load of the cubic to $O(N^4)$ method for a system  of 16 Si atoms.  Left: data generated by using fixed percentage errors in $\hat P$ of 0.1, 1, 10 and 20\% for  interpolation; 0.1, 1, 10, 30, and 50\% for naive Laplace; and 0.1, 1, 10, 30, 50, and 80\% for the windowed Laplace  for bulk Si.   Right: same for bulk MgO.  Fixed errors are set to be 0.1, 1, and 10\% for interpolation; 0.1, 1, 10, 30, and 70\% for naive Laplace; and 0.1, 1, 10, 20, and 40\% for windowed Laplace.}
\label{fig:inveps}
\end{figure}

\begin{figure}
\includegraphics[width=2.4in]{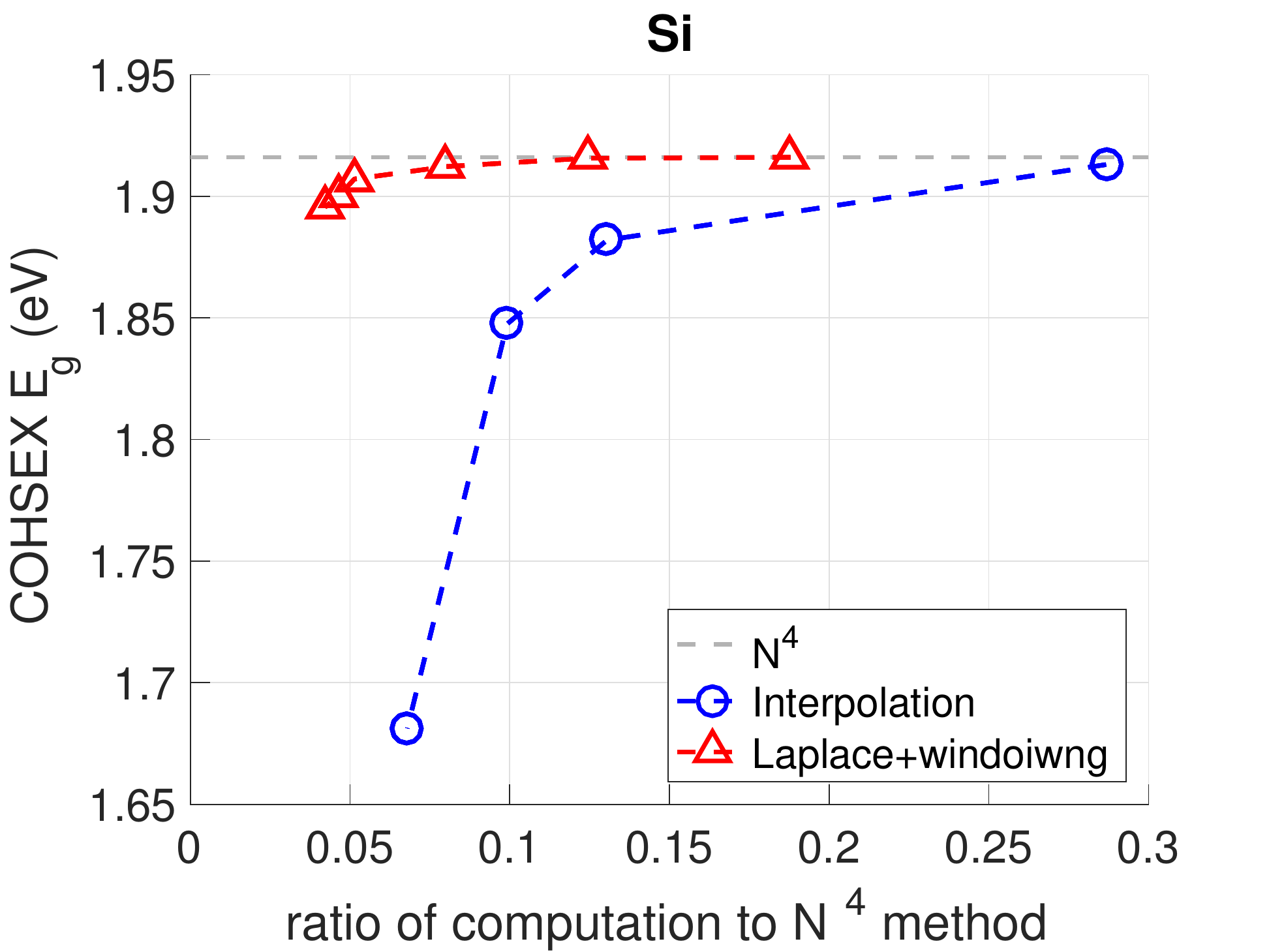}
\includegraphics[width=2.4in]{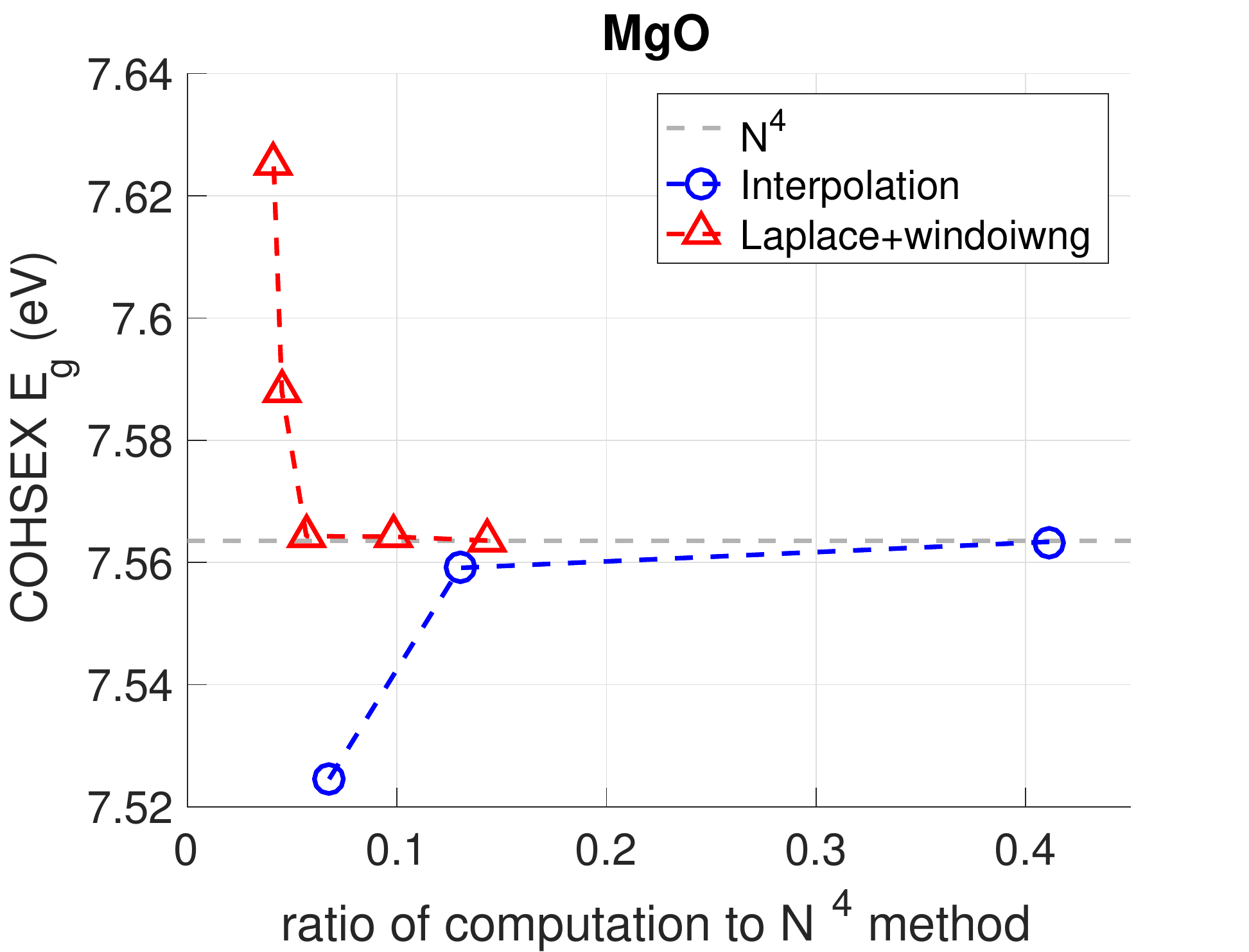}
\includegraphics[width=2.4in]{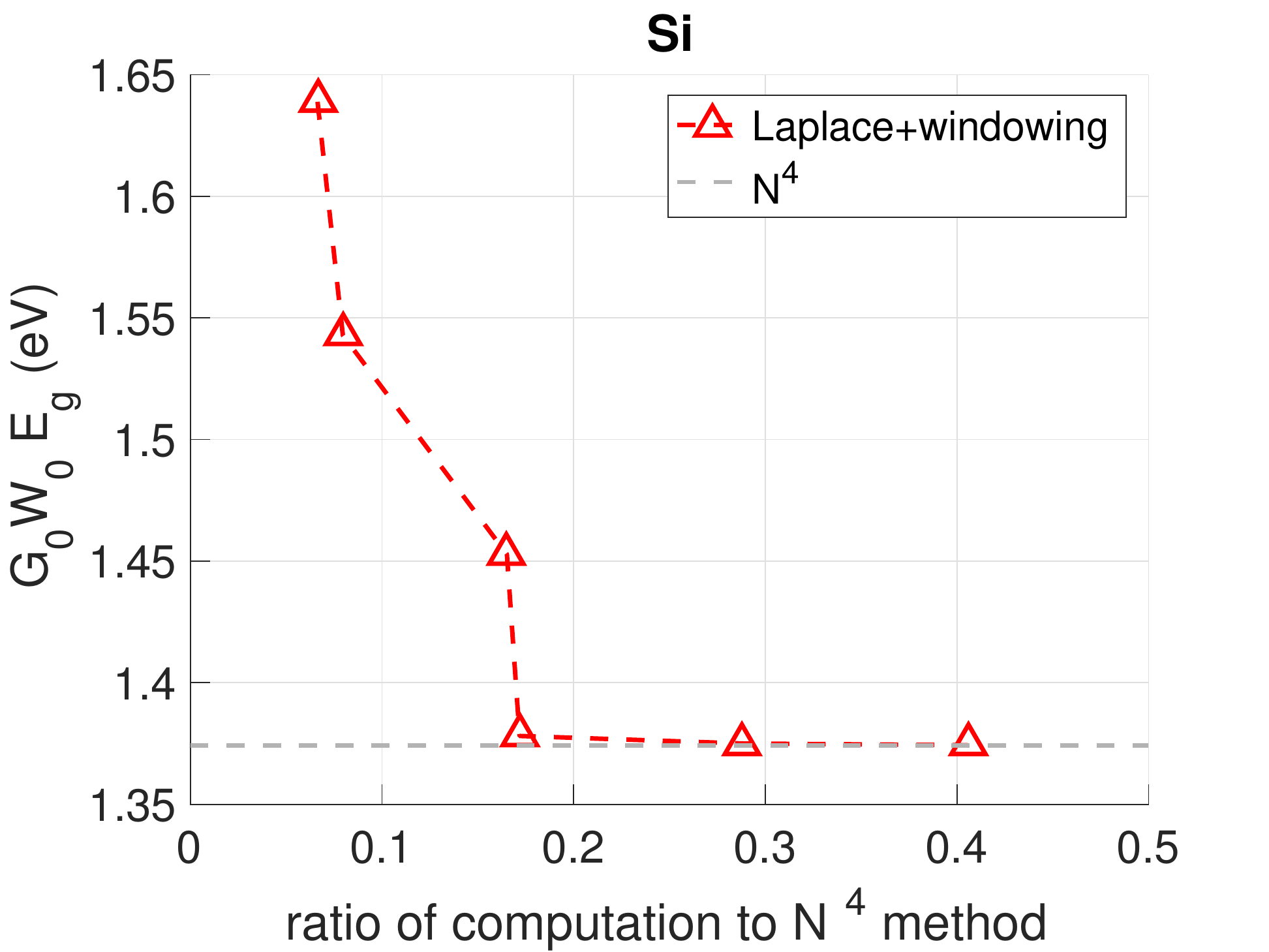}
\caption{Error of the bulk band gap ($\Gamma-X$ gap for Si and at $\Gamma$ for MgO) for different methods as a function of computational savings over the ``exact'' quartic method (horizontal dashed line).  All data are for a fixed system size of 16 atoms.  Same nomenclature and approach as Fig.~\ref{fig:inveps}.  The top  two figures are COHSEX approximation gaps and the bottom figure is the G$_0$W$_0$ Si band gap.}
\label{fig:bandgap}
\end{figure}

To evaluate the performance of our method, we chose two  materials: Si and MgO.  We run standard plane wave pseudopotential DFT calculations for both materials to describe the ground state and DFT band structure~\cite{supplement}.
Si is a prototypical covalent crystal with a moderate band gap (0.5 eV in DFT-LDA) while rocksalt MgO is an ionic crystal with a relatively large  gap (4.4 eV with LDA). 
We monitor the errors in two basic observables: the macroscopic optical dielectric constant $\epsilon_{\infty}$ and the band gap.  Figure~\ref{fig:inveps} shows the error in $\epsilon_{\infty}$ as a function of the computational savings achieved by our $N^3$ method compared to the $N^4$ method for a fixed system size of 16 atoms.  Each data point is generated by fixing a maximum error tolerance for $\hat P$ to derive parameters for energy windows and GL quadratures.  Then the error tolerance is varied to generate the plots.  Figure~\ref{fig:bandgap} shows data for the band gaps within the COHSEX approximation for the GW self-energy~\cite{hedin_new_1965}.

The windowed Laplace GL approach is the clear winner, especially for Si which has a much smaller band gap than MgO. The interpolation method works better for MgO than Si: the larger gap in MgO means that functions of energy are easier to interpolate.  For both materials, we achieve better than 0.1 eV accuracy of the band gap with at least an order of magnitude reduction in computation.  These results are for a fixed system size of $N=16$ atoms, so the savings improve linearly with the number of atoms for $N>16$.
% Figure - band gap

For a complete GW calculation, one must handle metallic systems and also compute the self-energy.  For metals, one replaces $1/[E_c-E_v]$ in Eq.~(\ref{eq:P}) by $[f(E_v)-f(E_c)]/[E_c-E_v]$ where $f(E)$ is a smoothed step function around the chemical potential $\mu$ (Fermi level)~\cite{fu_first-principles_1983,needs_total-energy_1986,gillan_calculation_1989} which leads to smooth behavior when $E_v=E_c=\mu$; only minor changes to our method are needed~\cite{supplement}.  Turning to the self-energy, if the poles of the screened interaction $W(\omega)_{r,r'}$ are at $\omega_p$ with residues $B_{r,r'}^p$, the dynamic (frequency-dependent) part of the GW self-energy is
\begin{equation}
\Sigma(\omega)^{dyn}_{r,r'} = 
\sum_{p,n}\frac{B^p_{r,r'}\psi_{rn}\psi_{r'n}^*}{\omega-\epsilon_n+sgn(\mu-\epsilon_n)\omega_p}\,.
\label{eq:sigma}
\end{equation}
 We can apply  windows-plus-quadrature to generate a cubic scaling method that delivers  $\Sigma^{dyn}(\omega)$ directly for real frequencies $\omega$~\cite{noteonsigma}.  We create two sets of windows for the two sets of energies $\{\omega-\epsilon_n\}$ and $\{\omega_p\}$ and write $\Sigma^{dyn}$ as a sum over window pairs as per Eq.~(\ref{eq:Pwindowed}) where each window pair has its own quadrature. Almost all the terms in Eq.~(\ref{eq:sigma}) can use the above Laplace with GL quadrature scheme with no modification since the denominator $x=\omega-\epsilon_n\pm\omega_p$ is finite and with fixed sign for two non-overlapping windows.  The difficulty is that, for overlapping windows, the denominator $x$ changes sign inside the energy windows so we can not use Eq.~(\ref{eq:laplace}). 
 We have created a Gaussian-type quadrature for the overlapping window cases~\cite{supplement} that delivers accurate results with small quadrature grids.  Figure~\ref{fig:bandgap} shows the method in action for the band gap of Si: high accuracy is possible with large computational savings compared to the $N^4$ method.

% Figure - scaling
\begin{figure}
\centering
\includegraphics[width=2.4in]{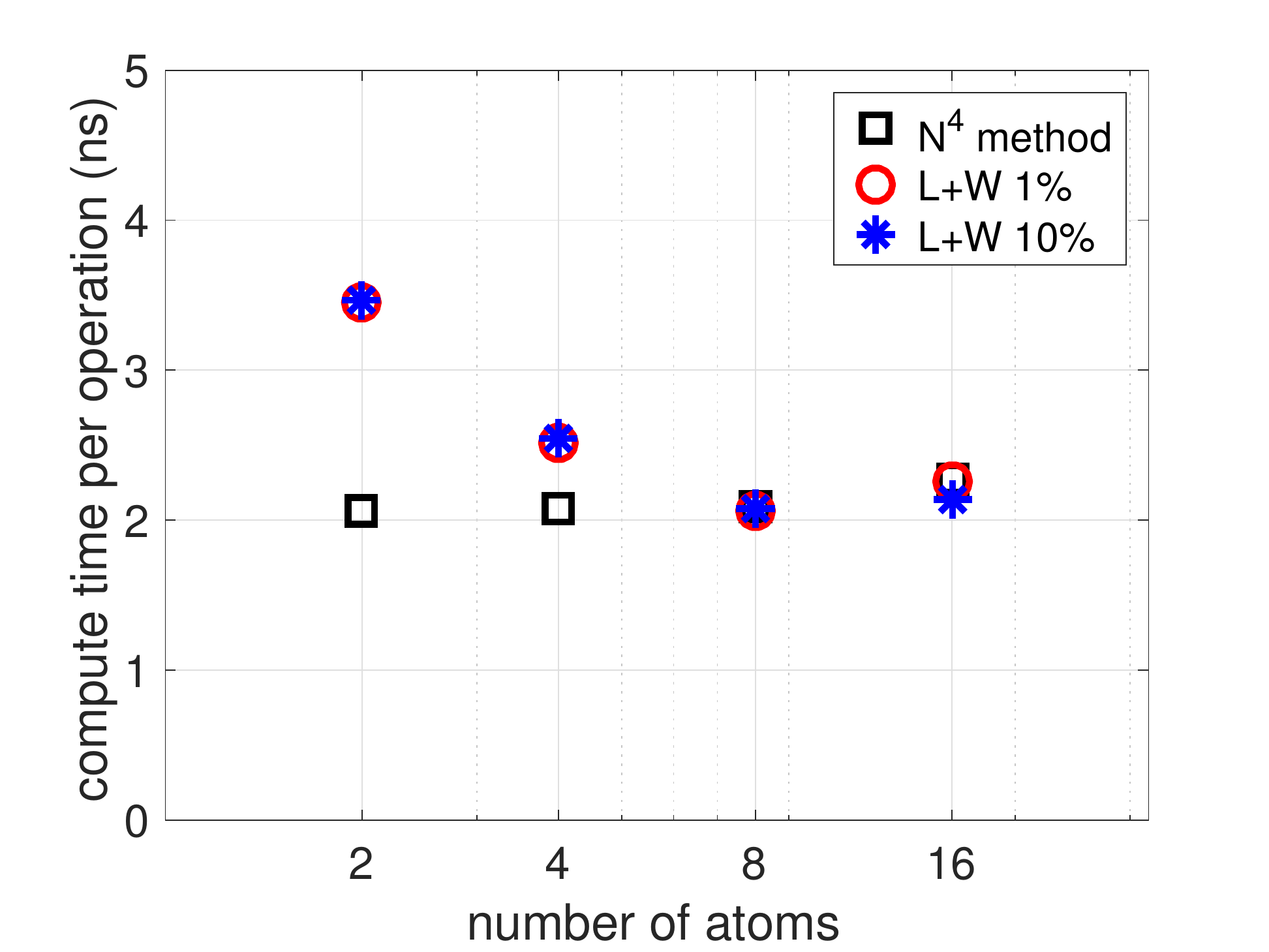}
\caption{Compute time per operation for evaluation of $P$. Black squares indicate the $N^4$ method, and red circles and blue asterisks indicate the $N^3$ Laplace windowed GL method (LW) with accuracy settings of 1\% and 10\% for $\hat{P}$. A serial linux computer is used.}
\label{fig:scaling}
\end{figure}
The final point is to verify the scaling of our method and to see if it confers any benefits compared to existing methods in practice.  To verify scaling, 
we time the $P$ calculation versus the number of atoms and show the compute time per operation in Figure~\ref{fig:scaling}: the number of operations are $N_v N_c N_r^2$ for the $N^4$ method and $\sum_{l,m}N_{GL}^{lm}(N_c^m+N_v^l)N_r^2$ for the windowed Laplace.   The essentially flat nature of the data shows that the algorithms scale as claimed~\cite{noteonlapalcefewatoms}. It is  exciting that all the compute times {\it per operation}  are very close to each other:  our $N^3$ method has a prefactor that is comparable to the $N^4$ method already for small systems, so we get a speedup even for small $N\gtrapprox10$.
A direct comparison between our windowed Laplace method and  accelerated $O(N^4)$ methods shows that our method is already competitive for small systems~\cite{supplement}.  Comparison to existing $O(N^3)$ methods also shows sizable improvements for our windowed Laplace approach~\cite{supplement}.

In summary, we have presented a real-space cubic-scaling sum-over-states method for GW calculations that works directly in frequency space and does not require analytic continuation from the imaginary to real axis. The method is already competitive with standard $N^4$ scaling methods for unit cells of 10-20 atoms as above and provides significant computational savings for desired band gap accuracies of 10-100 meV.  Finally, the method is straightforward to implement in a number of existing GW implementations using any basis set for which an efficient DVR can be constructed.

\begin{acknowledgments}
We thank Jack Deslippe and Gian-Marco Rignanese for helpful discussions. This work was supported by the NSF via grant ACI-1339804.
\end{acknowledgments}

% --------------------- Bibliography
%\bibliography{refs}

\begin{thebibliography}{37}
\bibitem [{not()}]{noteonsigma}%
  \BibitemOpen
  \href@noop {} {}\bibinfo {note} {To avoid excessive memory use, one can compute the large matrix $\Sigma(\omega)^{dyn}_{r,r'}$ for a fixed $\omega$ and then compute and only store the much smaller number of desired matrix elements $\langle n| \Sigma(\omega)|n'\rangle$ before moving to the next $\omega$ value.}\BibitemShut {Stop}%
\bibitem [{not()}]{noteonlapalcefewatoms}%
  \BibitemOpen
  \href@noop {} {}\bibinfo {note} {The windowed Laplace runs slightly slower
  per operation for few atoms which is due to inefficient caching and
  pipelining at small problem sizes.}\BibitemShut {Stop}%
\end{thebibliography}
%merlin.mbs apsrev4-1.bst 2010-07-25 4.21a (PWD, AO, DPC) hacked
%Control: key (0)
%Control: author (8) initials jnrlst
%Control: editor formatted (1) identically to author
%Control: production of article title (-1) disabled
%Control: page (0) single
%Control: year (1) truncated
%Control: production of eprint (0) enabled
%

\end{document}